# Flexoelectret: An Electret with Tunable Flexoelectric-like Response

*Xin Wen, Dongfan Li, Kai Tan, Qian Deng\* and Shengping Shen\**

X. Wen, K. Tan, Prof. Q. Deng, Prof. S. Shen.
State Key Laboratory for Strength and Vibration of Mechanical Structures, Xi'an Jiaotong University, Xi'an 710049, China
E-mail: tonydqian@mail.xjtu.edu.cn; sshen@mail.xjtu.edu.cn
D. Li
School of Science and Frontier Institute of Science and Technology, Xi'an Jiaotong University, Xi'an 710049, China

(Xin Wen and Dongfan Li contributed equally to this work.)

(Abstract) Because of the flexoelectric effect, dielectric materials usually polarize in response to a strain gradient. Soft materials are good candidates for developing large strain gradient because of their good deformability. However, they always suffer from lower flexoelectric coefficients compared to ceramics. In this work, a flexoelectric-like effect is introduced to enhance the effective flexoelectricity of a polydimethylsiloxane (PDMS) bar. The flexoelectric-like effect is realized by depositing a layer of net charges on the middle plane of the bar to form an electret. Experiments show that the enhancement of the flexoelectricity depends on the density of inserted net charges. It is found that a charged layer with surface potential of -5723V results in 100 times increase of the material's flexoelectric coefficient. We also show that the enhancement is proportional to the thickness of electrets. This work provides a new way of enhancing flexoelectricity in soft materials and further prompt the application of soft materials in electromechanical transducers.

Flexoelectricity is a widely observed property of dielectric materials (including crystals [1, 2], polymers [3], biomembranes [4], bones [5], liquid crystals [6], semiconductors [7], et al), which couples the strain gradient and the electric polarization. When a dielectric material is deformed non-uniformly, regardless of its initial crystalline symmetry, the strain gradient would break or change its spatial inversion symmetry and consequently induce a change of the polarization [8, 9]. In materials showing flexoelectricity, the ratio of the change of polarization to the strain gradient is defined as the flexoelectric coefficient. Since the strain gradient scales up with the decrease of the sample size, large strain gradients are usually found at small length scales. For this reason, several applications and/or ramifications of flexoelectricity at the nanoscale have been reported recently [10-19]. However, a macroscopic strain gradient is usually too small to produce relatively strong polarization unless the material has a very large flexoelectric coefficient [20-23].

To obtain large strain gradient at the macroscopic scale, a possible way is to use soft materials which are able to endure large strain before failure. So it is interesting to study flexoelectricity in polymers with much better deformability than ceramics [24, 25]. Other reasons for studying flexoelectricity in polymers lie in the facts that they are biocompatible, environmentally friendly, and suitable for applications in stretchable electronics. However, one of the biggest issues for flexoelectric polymers is their small flexoelectric coefficients. Chu and Salem have measured the flexoelectric coefficient for polymers to be $10^{-9} \sim 10^{-8} Cm^{-1}$ [3] which is several orders of magnitude smaller than that for ferroelectric materials reported by Ma and Cross ($\sim 10^{-6}\ Cm^{-1}$) [20,21]. Thus, in addition to the introduction of large strain, it is necessary to further enhance the flexoelectric coefficient of polymers in order to achieve strong flexoelectric effect.

In this work, other than increasing the intrinsic flexoelectric coefficient, we introduce a flexoelectric-like effect in electret materials. An electret is a dielectric material which contains quasi-permanent electrical charges. A well-known example for the application of electret materials is the ferroelectret (or piezoelectret), a cellular polymer film with charged air bubbles. Each charged air bubble may be viewed as a capacitor. When the material is deformed, the shape of air bubbles changes accordingly. In this way, the overall capacitance and polarization of the material are changed with its deformation. In other words, the material exhibits effective piezoelectricity. The strength of this effective piezoelectricity is found to be several tens times greater than that of the most used piezoelectric polymer, poly(vinylidene fluoride) (PVDF) [26, 27]. In recent years, ferroelectret has shown great potential in multiple areas such as energy harvesting [28, 29], sensing [30, 31], and self-powered electronics [32, 33], due to its high flexibility and effective piezoelectricity.

Motivated by ferroelectrets, here we propose a new design, flexoelectret, which shows effective flexoelectricity upon the application of non-uniform deformation. The structure of a flexoelectret is schematically drawn in **Fig. 1**(a). As shown in the figure, a layer of spatial charges are placed on the middle plane of a PDMS bar. The reason we choose PDMS instead of the commonly used flexoelectric polymer PVDF is that PDMS is much more deformable compared to PVDF. Larger deformation is helpful for creating large strain gradient, and thus positive to the flexoelectric effect. Because of the charge layer placed on the middle plane of the PDMS bar, the material is polarized and the induced polarizations above and below the charge layer have exactly the same magnitude $P_0$ but opposite directions (Fig. 1(b)). Thus, the overall polarization of the charged PDMS bar is zero. If we apply a uniform tension or compression to the bar, as shown in Fig. 1(c), its thickness and length would change in response to the mechanical loading. Consequently, the polarization would also change from $P_0$ to $P_1$. At this time, since the uniform deformation does not break the symmetry of the bar about its middle plane, the overall polarization is still zero. In other words, this charged PDMS bar shows no effective piezoelectricity. Actually, shifting the charge layer away from the middle plane does not contribute to the effective piezoelectricity since uniform deformation cannot change the net polarization of the sample. This point can be verified by Kacprzyk's theoretical work in which the effective piezoelectric coefficient of an electret is given by [34]:

$$d_{33}^{eff} = -q_0 \frac{\varepsilon_1 \varepsilon_2 h_1 h_2}{(\varepsilon_1 h_1 + \varepsilon_2 h_2)} \left(\frac{1}{Y_2} - \frac{1}{Y_1}\right) \tag{1}$$

where $q_0$ is the density of the charge layer, $\varepsilon_1$ and $\varepsilon_2$ respectively correspond to the permittivity value of the materials below and above the charge layer, $h_1$ and $h_2$ are the distance from the charge layer to the lower and upper surfaces of the bar, and $Y_1$, $Y_2$ denote the Young's modulus of the materials below and above the charge layer, respectively. This formula predicts the effective piezoelectricity of a double layered sample under uniform compression or tension. It is seen from the formula that the materials above and below the charge layer need to be different in order to produce non-zero effective piezoelectricity. For the model shown in Fig. 1(a), obviously $Y_1$ is equal to $Y_2$. So the effective piezoelectric coefficient $d_{33}^{eff}$ vanishes no matter whether $h_1$ is equal to $h_2$ or not.

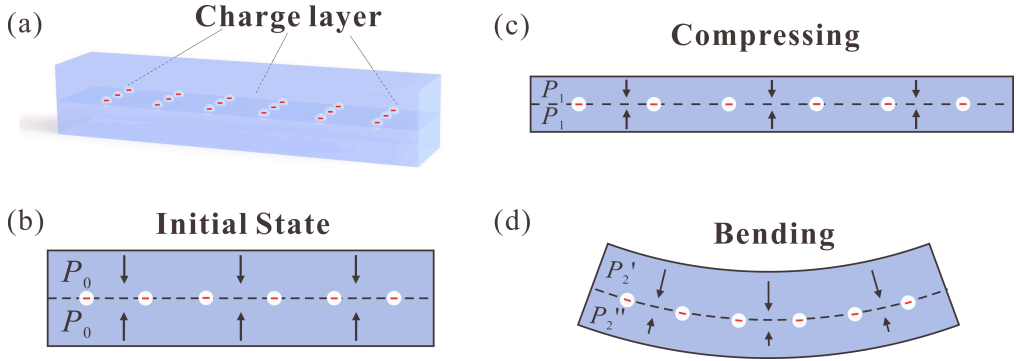

**FIG. 1.** Schematic illustration of the mechanism of the flexoelectret. Arrows in this figure represent polarization. Longer arrows correspond to larger polarization. (a) Structure of the flexoelectret. A layer of charges are deposited on the middle plane of a PDMS bar. (b) Initial state of the flexoelectret. The polarizations above and below the charge layer have the same magnitude $P_0$ but opposite directions. (c) Under uniform compression, the shape of the bar has changed but is still symmetric about the middle plane. The polarizations of two parts still have the same magnitude $P_1$ and opposite directions. (d) Under bending, symmetry of the bar is broken. The magnitudes of polarization $P_2'$ and $P_2''$ are no longer the same, which generates net polarization along the thickness direction.

While if subjected to a non-uniform loading, for example, a bending moment, the PDMS bar shown in Fig. 1(a) would no longer keep its symmetry. The upper and lower parts of the sample would experience totally different deformations. Upon a pure bending, the lower layer would expand while the upper layer would shrink. Thus, the strain across the thickness direction is non-uniform. Fig. 1(d) shows that, because of this biased deformation, the magnitudes of polarization $P_2^{'}$ and $P_2^{''}$, which respectively correspond to the polarization above and below the charge layer, may not cancel each other and there are net polarizations developed in the sample. Further bending would enhance the difference between the deformations of the upper and lower surfaces. Thus, the net polarization increases with the degree of the bending.

It is worthwhile to mention that such bending-induced redistribution of polarization commonly exists in lipid bilayer membranes and is proposed to be the mechanism of the flexoelectricity of biomembranes [4, 35-37]. The lipid bilayer consists of two layers of lipid molecules which are arranged in such a way that their hydrophilic heads are pointing outside and in contact with the environment while hydrophobic tails are pointing inside and isolated. As shown in **Fig. 2**(a), there are initial polarization in the biomembrane. If the membrane is flat and the environment on both sides are the same, there is no net polarization because the bilayer is symmetric about its middle plane. While, as we can see from Fig. 2(b), bending the membrane would break this symmetry and induce net polarization which is proportional to the membrane's curvature. In reality, no lipid bilayer is absolutely flat and there are also protein molecules embedded in the biomembrane. Thus, initially, the system is not symmetric and there are net polarizations. Further bending the membrane would change the degree of the asymmetry and thus lead to a further change of its net polarization.

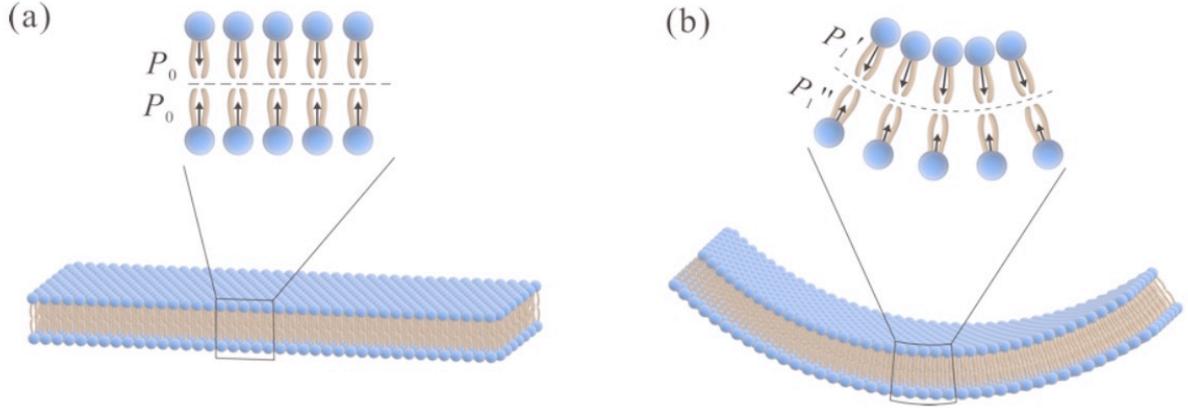

**FIG. 2.** Structure of a lipid bilayer membrane. Arrows represent the average polarization of the lipid molecules. Longer arrows correspond to larger polarization. (a) A flat lipid bilayer membrane, which is symmetric about its middle plane. (b) A curved lipid bilayer membrane. The symmetry of membrane is broken, which leads to net polarization.

Since all dielectric materials show flexoelectricity, we firstly performed a three-point bending test on pure PDMS without net charges to check how the material's intrinsic flexoelectricity would affect its electromechanical behavior. **Fig. 3(a)** shows the experimental setup for flexoelectric measurement. The test sample is a PDMS bar with the size of 100mm× 15mm×10mm and coated with compliant electrodes (liquid metal) on its upper and lower surfaces. A load machine applied a sinusoidal oscillatory force to the center point of the bar's upper surface. The frequency of the oscillatory force was set to 1 Hz, much lower than the bar's natural frequency. Thus, the inertial effect can be simply ignored in theoretical analysis. Two supporting edges were separated by $L$=80mm. The vertical displacement of the center point of the bar's upper surface $\delta$ was recorded (peak value was equal to 1.5mm). The bending-induced polarization charges $Q$ were detected using a charge amplifier and an oscilloscope. Then the intrinsic transverse flexoelectric coefficient $\mu_{13}$ can be calculated according to

$$\overline{P_3} = \mu_{13} \overline{\frac{\partial \varepsilon_{13}}{\partial x_3}} \tag{2}$$

and

$$\overline{P_3} = \frac{Q}{A} \tag{3}$$

where $A$ is the area of the electrode, and $\overline{\dfrac{\partial \varepsilon_{13}}{\partial x_3}}$ is the average strain gradient across the thickness direction, which can be calculated by finite element method (FEM) based on measured $L$ and $\delta$ (see the Supplemental Material [38]). In order to eliminate electromagnetic interferences from the environment, we filtered the data and extracted the signal at 1 Hz. In Fig. 3(b), the red dash line and blue solid line depict the variation of charge output $Q$ and vertical displacement $\delta$ for the case of pure PDMS bar. We found through calculations that intrinsic flexoelectric coefficient $\mu_{13}$ of PDMS is $5.3\times 10^{-10} Cm^{-1}$, much lower than the measured value for PVDF ($\sim 10^{-8} Cm^{-1}$).

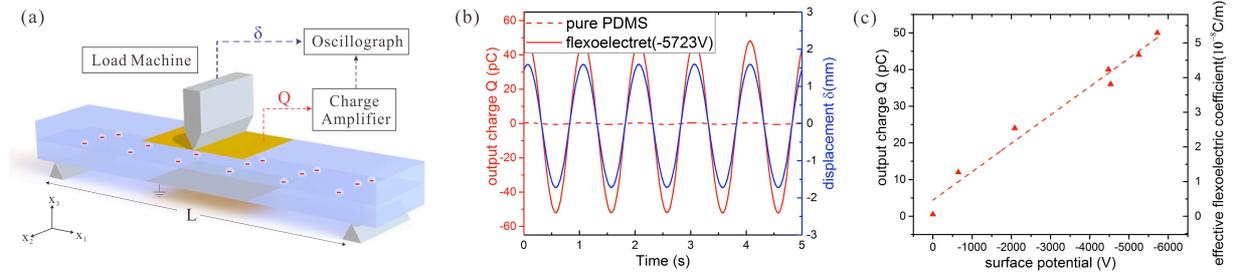

**FIG. 3.** Experimental setup and results. (a) Experimental setup for flexoelectric measurement. (b) Output charge $Q$ from the upper electrode and vertical displacement of the center point of the sample, corresponding to the red and blue lines respectively. (c) Output charge $Q$ and the effective flexoelectric coefficient $\mu_{13}^{eff}$ as a function of the surface potential of the charge layer.

Then we made a flexoelectret by placing a piece of charged thin film with the density of $q_0$ on the middle plane of the PDMS bar (see the Supplemental Material [38]) and repeated the above mentioned three-point bending test. Note that measuring charge density directly is difficult. So we use surface potential to characterize the charge density of the charged film. Larger surface potential corresponds to larger charge density and these two quantities have a roughly linear relationship to each other.[33] The experimental result is also given by Fig. 3(b). In Fig. 3(b), the red solid line depicts the variation of the output charge $Q$ with respect to time in response to the same displacement $\delta$ as shown by the blue solid line. Comparing the red dash line and the red solid line in Fig. 3(b), we found a strong enhancement of the electromechanical coupling of the flexoelectret over pure PDMS. Following exactly the same process for calculating the intrinsic flexoelectric coefficient above, we measured the effective flexoelectric coefficient of the flexoelectret $\mu_{13}^{\text{eff}}$ to be $5.3\times10^{-8}\text{Cm}^{-1}$ when the surface potential of the charged film is -5723V. This value is two orders of magnitude higher than that of pure PDMS. Although this flexoelectic coefficient is at the same order of magnitude as that for PVDF, much larger deformability of PDMS over PVDF allows more significant flexoelectric effect in the former if testing at the same length scale. Note that this effective flexoelectricity is caused neither by the non-centrosymmetric distortion of the microstructure (the origination of intrinsic flexoelectricity) nor by the uniform change of the capacitance (the origination of piezoelectret effect). The flexoelectric-like effect observed here stems from the redistribution of pre-existing polarizations under non-uniform deformation. Note that the pre-existing polarization mentioned here is caused by the net charges placed on the middle plane. If there is no pre-existing polarization, the flexoelectric-like effect would vanish. We propose here that larger pre-existing polarization should lead to stronger flexoelectric-like effect. To confirm this point, the same three-point bending test was conducted on electrets with different densities of charges on the middle plane. It is found from Fig. 3(c) that the output charge increases almost linearly with respect to the charge density of the middle plane.

To explore the mechanism of the above phenomenon, the problem was analyzed using FEM (COMSOL Multiphysics 5.2). **Fig. 4(a)** illustrates the schematic diagram of established model with the same size and material parameters as experimental specimen (see the Supplemental Material [38]). The red solid line represents a layer of charge embedded in the interface between two layers of same dielectric material. Note that the polarization and electric displacement fields above and below the interface have opposite directions, to accurately distinguish the difference between their magnitudes, we plotted absolute value of the polarization and electric displacement in Fig. 4(b), (c), (e), and (f). Fig. 4(b) shows that the pre-existing polarization distribution is uniform, which means the overall polarization is zero. When subjected to uniform pressure, both the shape and the polarization distribution of the sample changed accordingly. However, as shown in Fig. 4(c), the overall polarization is still zero. This is because that the symmetry of the model has not been broken by the uniform strain.

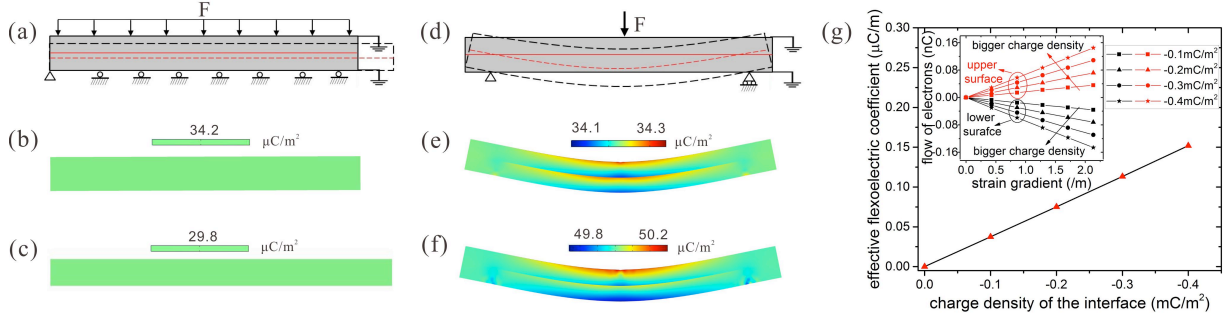

**FIG. 4.** Schematic illustration of the FEM model of flexoelectrets and calculated results. The polarization and electric displacement distribution are all absolute values. Dash lines represents the deformed model. (a) Model under uniform pressure. (b) Pre-existing polarization distribution in the initial state, which is uniform. (c) Polarization distribution under uniform compression, which is still uniform. (d) Model under three-pointing bending. (e,f) Polarization and electric displacement distribution under bending respectively, showing a strong asymmetry. (g) Calculated effective flexoelectric coefficient as a function of charge density. The inset shows the dependence of the electronic flow on the strain gradient and charge density.

Then we bended the bar by applying concentric force $F$ to the middle point of its upper boundary and fixing its lower left and lower right ends (Fig. 4(d)). Fig. 4(e) and (f) show the redistribution of polarization and electric displacement, respectively. It can be seen clearly that the symmetry of the model was broken due to the bending deformation. Fig. 4(g) (inset) shows the charges flowing from the lower surface to the upper surface (see the Supplemental Material [38]). Obviously, further bending the sample and larger charge density both lead to more charges flow. Then we obtained the effective flexoelectric coefficient $\mu_{13}^{\text{eff}}$ using equation (2). It can be seen from Fig. 4(g) that $\mu_{13}^{\text{eff}}$ varies linearly with respect to the charge density, which is well consistent with our experimental measurement shown in Fig. 3(c).

We further found that the effective flexoelectric coefficient is also related to the thickness of flexoelectrets. We used specimens with different thickness (2mm, 5mm, and 10mm) but same length and width for flexoelectric testing. The measured (calculated) coefficients $\mu_{13}^{\text{eff}}$ are divided by surface potential $V$ (charge density $q$) to eliminate the influence of the charge density. **Fig. 5** shows that the normalized effective flexoelectricity is proportional to sample thickness. Actually, in a bent flexoelectret with certain curvature (or strain gradient), the effective flexoelectricity is mainly contributed by the parts away from the middle plane. This can be seen clearly from Fig. 4(f), where the maximum and minimum values of the electric displacement are found at the upper and lower surfaces, respectively. If the sample's thickness is decreased while keeping the curvature unchanged, according to Fig. 4(f), the absolute values for the electric displacement on the upper and lower surfaces would drop accordingly. So the effective flexoelectricity also drops. On the other hand, increase the thickness would lead to the increase of the electric displacement on the upper and lower surfaces, which corresponds to a further enhancement of the effective flexoelectricity.

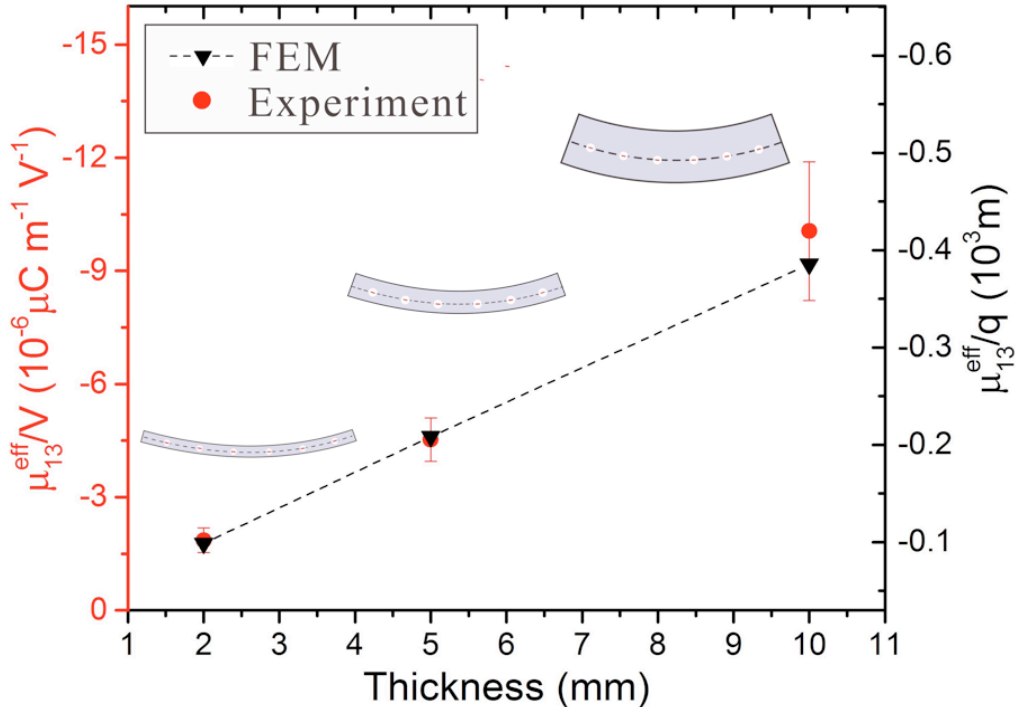

**FIG. 5.** Normalized effective flexoelectricity as a function of sample thickness. Both experimental results (filled red circles) and FEM results (filled black triangles) show a linear relationship between flexoelectricity and thickness.

The strength of the flexoelectric effect in materials depends on their flexoelectric coefficients as well as the strain gradient that can be developed in them. In this work, large strain gradients are realized through the using of highly deformable materials and, at the same time, the flexoelectric coefficient is enhanced by an electret based flexoelectric-like phenomenon which stems from the interaction between non-uniform deformations and the initial polarization field. Specifically, a PDMS bar is used for the flexoelectret, and the initial polarization field is introduced by a layer of net charge deposited in its middle plane. Experiments indicate that this charged PDMS bar has an effective flexoelectric coefficient which is 100 times of its intrinsic flexoelectric coefficient. It is also found both experimentally and through FEM that the effective flexoelectric coefficient $\mu_{13}^{\text{eff}}$ of this flexoelectret linearly depends on both the sample thickness and the density of net charges deposited on its middle plane. This work provides an example of the interaction between non-uniform deformations and the polarization field, introduces a new way of enhancing flexoelectric coefficients in soft materials, and also shows a novel application of electrets. Since current flexoelectric energy harvesting is usually based on nanobeams with extremely high resonant frequencies, the flexoelectret introduced here may have promising applications in low frequency flexoelectric energy harvesting.

**Acknowledgements**
We gratefully acknowledge the support from the National Natural Science Foundation of China (Grant Nos. 11632014, 11672222, and 11372238), the 111 Project (B18040), the Chang Jiang Scholar Program, and numerous helpful discussions with Dr. Guanghao Lu and Dr. Wenfeng Liu.